\documentclass[aps,nofootinbib,11pt,
]{revtex4}
\usepackage{amsmath}
\usepackage{amsfonts}
\usepackage{amssymb}

\usepackage{color}

\newcommand{\omits}[1]{}

\begin{document}


\title{Spinor Particle Creation in Near Extremal Reissner-Nordstr{\"o}m Black Holes}

\author{Chiang-Mei Chen} \email{cmchen@phy.ncu.edu.tw}
\affiliation{Department of Physics and Center for Mathematics and Theoretical Physics, National Central University, Chungli 320, Taiwan}

\author{Jia-Rui Sun} \email{jrsun@ecust.edu.cn}
\affiliation{Department of Physics and Institute of Modern Physics, East China University of Science and Technology, Shanghai 200237, China}

\author{Fu-Yi Tang} \email{foue.tang@gmail.com}
\affiliation{Department of Physics, National Central University, Chungli 320, Taiwan}

\author{Ping-Yen Tsai} \email{a3675531@hotmail.com}
\affiliation{Department of Physics, National Central University, Chungli 320, Taiwan}

\date{\today}


\begin{abstract}
The pair production of spinor particles, which can be captured by the solution of the Dirac equation with an appropriate boundary condition, in charged black holes is investigated. We obtain the closed form of the production rate in the near extremal limit of the Reissner-Nordstr{\"o}m black holes. The cosmic censorship conjecture is seemingly guaranteed by the pair production process. Moreover, the absorption cross section ratio and retarded Green's functions of the spinor fields calculated from the gravity side match well with those of spinor operators in the dual CFT side.
\end{abstract}


\maketitle
\tableofcontents

\section{Introduction}
The spontaneous pair production in which the virtual particles and antiparticles from vacuum fluctuations are separated, by various mechanisms, to be real pairs is essentially a significant quantum phenomenon. In particular, the pair production occurring in charged black holes mixtures two independent processes, namely the Schwinger mechanism by an electromagnetic force~\cite{Schwinger:1951nm} and the Hawking radiation by a tunneling through horizon~\cite{Parikh:1999mf}. There are numerous studies of pair production in the Reissner-Nordstr{\"o}m (RN) black holes in the literature~\cite{Zaumen:1974, Carter:1974yx, Damour:1974qv, Gibbons:1975kk, Page:1977um, Hiscock:1990ex, Khriplovich:1999gm, Gabriel:2000mg, Belgiorno:2007va, Belgiorno:2008mx, Belgiorno:2009pq}.

In our previous study~\cite{Chen:2012zn}, we considered the pair production of scalar particles created from a particular background---the near horizon region of near extremal RN black holes. The consideration for only taking into account the effect at the near horizon is inspired by an intuitive expectation that the pair production should mainly occur at this region which contains the causal boundary for the Hawking radiation and the dominated electric field for the Schwinger mechanism. The final results indeed confirm such anticipation. In addition, the primary motivation for our study is to understand the holographic dual interpretation for the pair production process in black holes in the context of the AdS/CFT correspondence. The holographic description of black holes has been intensively studied after the leading work on Kerr/CFT correspondence~\cite{Guica:2008mu}, see, e.g., the recent review~\cite{Compere:2012jk} and the references therein. The preliminary information is usually derived from the study on the near extremal black holes. For example, in the near extremal RN black holes the near horizon geometry has a particular AdS$_2 \times S^2$ structure which allows to verify the holographic duality with the knowledge of AdS/CFT correspondence~\cite{Maldacena:1997re, Gubser:1998bc, Witten:1998qj}.

In the present paper, we extend the previous study on scalar particles to explore the spinor pair production. We firstly solve the Dirac equation for a probe massive charged spinor field in the near horizon region of the near extremal RN black holes. Due to the spherical symmetry, the spinor field can be expanded by the spherical spinors~\cite{Th92}, then the Dirac equation can be separated and reduces to two first order coupled ordinary differential equations for two radial functions. Furthermore, these two equations can be transformed into the well-know hypergeometric equations. Consequently, the exact solutions are obtained in terms of the hypergeometric functions. From the behaviors of spinor field at the asymptotic (outer boundary) and horizon (inner boundary) we can identify the corresponding ingoing and outgoing modes on the boundaries. There are two equivalent boundary conditions to capture the pair production process~\cite{Chen:2012zn}. Here we will enforce the ``particle viewpoint'' boundary condition by imposing vanishing ingoing flux at outer boundary. The other three fluxes has intuitive physical interpretation from particle viewpoint: the outgoing flux at outer boundary representing the pair produced particles, and the outgoing and ingoing fluxes at inner boundary describing the virtual and re-annihilated particles respectively. Finally the physical quantities of pair production, i.e. the Bogoliubov coefficients (vacuum persistence and mean number of pairs) and the absorption cross section ratio, can be derived from the ratios of these three fluxes~\cite{Kim:2008yt, Kim:2009pg}. Moreover, the pair production can occur only when the spinor field generated at the horizon can propagate to the infinity. The existence condition, in both scalar and spinor fields, is identical with the violation of Breitenlohner-Freedman (BF) bound~\cite{Breitenlohner:1982jf, Breitenlohner:1982bm} in AdS$_2$. Thus, the pair production indeed corresponds to the instability of probe field (or perturbations) in AdS$_2$ (or effectively AdS$_3$ spacetime based on the RN/CFT correspondence~\cite{Chen:2009ht, Chen:2010bsa, Chen:2010as, Chen:2010yu, Chen:2010ywa, Chen:2011gz}). Consequently, the conformal weights of the spinor operator dual to the charged spinor field are complex indicating an instability in the dual CFT$_2$. It is worth to note that the existence of pair production strongly requires the charge of produced spinor particle should be bigger than its mass. By the charge and energy conservations\footnote{This argument is not completely rigorous since we only consider the near horizon region and the gravitational backreaction is neglected in the probe limit. In addition, it was claimed in~\cite{Richartz:2011vf} that the cosmic censorship conjecture for the near extremal RN black holes can be violated by quantum tunneling of spin-$\frac12$ charged particles.}, the black holes in the pair production process should lose charge more than mass, and this property ensures the cosmic censorship conjecture. In addition, using the RN/CFT correspondence, we show that the absorption cross section ratio and the retarded Green's function of the bulk spinor field computed from the gravity side match well with those of the spinor operators in the dual CFT$_2$, and thus the pair production ratio also can be understood holographically. For other previous studies on Schwinger pair creation in AdS or dS spacetime backgrounds, see for example~\cite{Pioline:2005pf, Kim:2008xv}, and see~\cite{Sato:2013dwa, Sato:2013hyw, Fischler:2014ama} for other studies on the dual CFT description of the Schwinger effect.

The rest parts of the paper is organized as follows: We analytically solve the Dirac equation for charged spinor fields in the near horizon near extreme RN black hole background in Section~\ref{sect:spinorRN}; Then the pair production rate and absorption cross section ratio of the charged spinor are obtained by choosing appropriate boundary conditions in Section~\ref{sect:spinorPC}; In~\ref{sect:holo}, the holographic description of the spinor absorption cross section ratio, the retarded Green's function and the pair production ratio are analyzed based on the RN/CFT correspondence; We drawn the the conclusions and discussions in Section~\ref{sect:conclusion}. Some useful properties of spherical spinors and hypergeometric function are listed in appendixes~\ref{ASS} and~\ref{AHF}.

\section{Spinor Field in the RN Black Holes}\label{sect:spinorRN}

\subsection{Dirac Equation}
The Dirac equation for a massive and charged spinor field, $\Psi$, in a curved spacetime coupled with an electromagnetic field, $A_\mu$, is given by
\begin{equation} \label{Dirac}
\left[ \gamma^a e_a{}^\mu (\partial_\mu + \Gamma_\mu - i q A_\mu) + m \right] \Psi = 0,
\end{equation}
where $m$ and $q$ denote the mass and charge of the spinor field. The tetrad $e_a{}^\mu$ and $\Gamma$ one-form represent the curved spacetime effects in the light of the metric $g_{\mu\nu}$ and connection one-form $\omega_{ab}$ which are defined as
\begin{equation}
\eta_{ab} = e_a{}^\mu e_b{}^\nu g_{\mu\nu}, \qquad \Gamma = \Gamma_\mu dx^\mu = \frac18 \left[ \gamma^a, \gamma^b \right] \omega_{ab} = \frac14 \gamma^a \gamma^b \omega_{ab}.
\end{equation}
In this paper, the Greek and Latin letters symbolize the coordinate and frame indices respectively, and the convention for the gamma matrices is given in the Appendix~\ref{ASS}.

For a spherically symmetric spacetime coupled with an electric field, the metric and gauge potential generally are characterized by three radial dependent functions, $f(r), \rho(r)$, and $\phi(r)$, as
\begin{equation}
ds^2 = - f(r) dt^2 + \frac{dr^2}{f(r)} + \rho^2(r) d\Omega_2^2, \qquad A = \phi(r) dt,
\end{equation}
and the explicit expressions of these functions depend on the exact solution. In this paper, we will consider the near horizon solution of near extremal RN black holes and the corresponding functions are~\cite{Chen:2012zn}
\begin{equation} \label{ERN}
f = \frac{r^2 - B^2}{Q^2}, \qquad \rho = Q, \qquad \phi = - \frac{r}{Q}.
\end{equation}
The Dirac equation can be formulated in different choices of frame in associate with different representation of spinor field. In the follows we are going to solve the Dirac equation in the rotation frame, $\hat {\boldsymbol{e}}^a := e^a{}_\mu dx^\mu$, defined by the tetrad
\begin{equation}
e^0{}_t = \sqrt{f}, \qquad e^1{}_r = \frac1{\sqrt{f}}, \qquad e^2{}_\theta = \rho, \qquad e^3{}_\varphi = \rho \sin\theta.
\end{equation}
Straightforwardly, we can compute the following essential $\Gamma$-term
\begin{equation}
\gamma^a e_a{}^\mu \Gamma_\mu = \gamma^1 \left( \frac{\sqrt{f}\, \rho'}{\rho} + \frac{f'}{4 \sqrt{f}} \right) + \gamma^2 \frac{\cot\theta}{2 \rho},
\end{equation}
and therefore the Dirac equation in spherically symmetric spacetimes explicitly reduces to
\begin{equation}
\left[ \gamma^0 \frac1{\sqrt{f}} \left( \partial_t - i q \phi \right) + \gamma^1 \sqrt{f} \left( \partial_r + \frac{\rho'}{\rho} + \frac{f'}{4f} \right) + \gamma^2 \frac1{\rho} \left( \partial_\theta + \frac{\cot\theta}{2} \right) + \gamma^3 \frac1{\rho \sin\theta} \partial_\varphi + m \right] \Psi = 0.
\end{equation}
Moreover, the above equation can be further simplified, basically absorbing those connection terms as a rescaling of spinor field, according to the following definition
\begin{equation}
\Psi = (- \det g \, g^{rr})^{-\frac14} \, \Phi = \rho^{-1} \, \sin^{-\frac12}\theta \, f^{- \frac14} \, \Phi.
\end{equation}
Finally the considerable version of the Dirac equation is
\begin{equation}
\left[ \gamma^0 \frac1{\sqrt{f}} \left( \partial_t - i q \phi \right) + \gamma^1 \sqrt{f} \, \partial_r + \gamma^2 \frac1{\rho} \, \partial_\theta + \gamma^3 \frac1{\rho \sin\theta} \, \partial_\varphi + m \right] \Phi = 0.
\end{equation}

\subsection{Exact Solution}
The spinor field in spherically symmetric spacetimes can be expanded in terms of the orthonormal spherical spinors~\cite{Th92}, $\Phi^\pm_{\kappa, n}$, whose components are defined by spherical harmonics. The parameter $\kappa = \mp (j + 1/2)$ is specified by the angular momentum quantum number $j$ (half-integer) and $n$ is its projection. The spherical spinors have desirable properties with respect to gamma matrices $\gamma^0, \gamma^1$ and the operator $\boldsymbol{\cal K} = \gamma^2 \partial_\theta + \gamma^3 \sin^{-1}\theta \partial_\varphi$ which allow to separate the Dirac equation. The useful properties of the spherical spinors are summarized in the Appendix~\ref{ASS}. Hence, we impose the following ansatz of the spinor field
\begin{equation}
\Phi(t, r, \theta, \varphi) = \mathrm{e}^{-i \omega t} \left[ R_+(r) \Phi^+_{\kappa, n}(\theta, \varphi) + R_-(r) \Phi^-_{\kappa, n}(\theta, \varphi) \right],
\end{equation}
and then the Dirac equation, by using the properties~(\ref{g01K}), reduces to two first order coupled equations for two radial functions $R_\pm(r)$ as
\begin{equation}
\left( \sqrt{f} \partial_r \pm \frac{\kappa}{\rho} \right) R_\pm - i \left( \frac{\omega + q \phi}{\sqrt{f}} \pm m \right) R_\mp = 0.
\end{equation}
Technically, it is more proper to rewrite the equations such that the terms including function $f$ be combined together. This can be achieved by introducing new functions ${\cal R}_\pm = R_+ \pm R_-$, and then the first order equations are transformed to
\begin{equation}
\left( \sqrt{f} \partial_r \mp i \frac{\omega + q \phi}{\sqrt{f}} \right) {\cal R}_\pm + \left( \frac{\kappa}{\rho} \pm i m \right) {\cal R}_\mp = 0.
\end{equation}

For our considered background~(\ref{ERN}) of the near extremal RN black holes, the equations to be solved are
\begin{equation}
\frac{\sqrt{r^2 - B^2}}{Q} \partial_r {\cal R}_\pm \mp i \frac{\omega Q - q r}{\sqrt{r^2 - B^2}} {\cal R}_\pm + \left( \frac{\kappa}{Q} \pm i m \right) {\cal R}_\mp = 0,
\end{equation}
in which the last coefficient is simply a constant. These two equations can be further expressed into a more desirable form in terms of the new radial coordinate $z$,
\begin{equation}
z = \frac{r + B}{2 B},
\end{equation}
and two rescaled radial functions ${\cal \tilde R}_\pm$,
\begin{equation}
{\cal \tilde R}_\pm = \Sigma^{\pm 1} {\cal R}_\pm, \qquad \Sigma = (2B)^{i a} z^{i \frac{\tilde a + a}2} (z - 1)^{-i \frac{\tilde a - a}2},
\end{equation}
where the parameters $a, \tilde a$ and $b$ ($b$ is defined for later convenience) are
\begin{equation} \label{aab}
a := q Q, \qquad \tilde a := \frac{\omega Q^2}{B}, \qquad b := \sqrt{(q^2 - m^2) Q^2 - \kappa^2}.
\end{equation}
Interestingly these three parameters are almost identical to the analogous parameters appearing in the scalar field pair production~\cite{Chen:2012zn} except the term $\kappa^2 = (j + 1/2)^2$ with half-integer $j$ in parameter $b$ being replaced by $(l + 1/2)^2$ with integer $l$. Finally, the first order coupled equations reduce to
\begin{equation} \label{1EQtR}
\Sigma^{\mp 2} \partial_z {\cal \tilde R}_\pm + \frac1{\sqrt{z (z - 1)}} \left( \kappa \pm i m Q \right) {\cal \tilde R}_\mp = 0,
\end{equation}
and they can be simply decoupled as two second order equations
\begin{equation}
z (1 - z) \partial_z^2 {\cal \tilde R}_\pm + \left[ \frac12 \mp i \left( q Q + \frac{\omega Q^2}{B} \right) - \left( 1 \mp i 2 q Q \right) z \right] \partial_z {\cal \tilde R}_\pm + (\kappa^2 + m^2 Q^2) {\cal \tilde R}_\pm = 0.
\end{equation}
These two decoupled equations are just the hypergeometric equation~(\ref{hgeq}) with parameters
\begin{equation}
\alpha_\pm = i (b \mp a), \qquad \beta_\pm = - i (b \pm a), \qquad \gamma_\pm = \frac12 \mp i (\tilde a + a).
\end{equation}
Thus the corresponding solutions are generically two independent hypergeometric functions, see Appendix~\ref{AHF}. However, the actual solutions for functions ${\cal \tilde R}_\pm$ must satisfy the first order coupled equations~(\ref{1EQtR}) which should give constraints on integration constants. Therefore, there is only one free integration constant for each radial function.

Lets firstly show how to obtain the solutions around the point $z = 0$ in details. A simple way to construct the general solutions, suppose ${\cal \tilde R}_\pm = {\cal \tilde R}_\pm^{(1)} + {\cal \tilde R}_\pm^{(2)}$, consistent with the first coupled equations is to chose the first ``half'' part of solution in~(\ref{Fz1})
\begin{equation} \label{Rpm1}
{\cal \tilde R}_\pm^{(1)} = C_{\pm} \, F\left( i (b \mp a), - i (b \pm a); \frac12 \mp i (\tilde a + a); z \right),
\end{equation}
and then use the equations~(\ref{1EQtR}) to determine the complementary part, ${\cal \tilde R}_\pm^{(2)}$, as
\begin{equation} \label{Rpm2}
{\cal \tilde R}_\pm^{(2)} = \bar C_\pm z^{\frac12 \pm i (\tilde a + a)} \, F\left( \frac12 - i (b \mp \tilde a), \frac12 + i (b \pm \tilde a); \frac32 \pm i (\tilde a + a); z \right),
\end{equation}
where the coefficients $\bar C_\pm$ are determined by $C_\pm$
\begin{equation} \label{barC}
\bar C_\pm = - (-1)^{-\frac12 \pm i (\tilde a - a)} C_\mp \, \frac{\kappa \pm i m Q}{\frac12 \pm i (\tilde a + a)} (2B)^{\pm i 2 a}.
\end{equation}
One can straightforwardly check that ${\cal \tilde R}_\pm^{(2)}$ will reproduce the their complementary part ${\cal \tilde R}_\pm^{(1)}$ via the equations~(\ref{1EQtR}).

\subsection{Asymptotic and Near Horizon Behaviors}
In order to obtain the ingoing and outgoing fluxes at both inner and outer boundaries, we need the asymptotic and near horizon behaviors of the spinor field solution. For analyzing the asymptotic behaviors at $z = \infty$ ($r = \infty$), it is convenient to express the solution around the point $z = \infty$. Following the similar approach, we firstly chose ``half'' solution, a convenient choice is the first part in~(\ref{Fz8}) for ${\cal \tilde R}_+^{\infty (1)}$ and second part in~(\ref{Fz8}) for ${\cal \tilde R}_-^{\infty (1)}$, as
\begin{equation}
{\cal \tilde R}_\pm^{\infty (1)} = C_\pm^\infty \, z^{\mp i (b - a)} \, F\left( \pm i (b - a), \frac12 \pm i (b + \tilde a); 1 \pm i 2 b; \frac1{z} \right),
\end{equation}
and the complementary half is determined by the first order equations~(\ref{1EQtR})
\begin{equation}
{\cal \tilde R}_\pm^{\infty (2)} = \bar C_\pm^\infty \, z^{\pm i (b + a)} \, F\left( \mp i (b + a), \frac12 \mp i(b - \tilde a); 1 \mp i 2 b; \frac1{z} \right),
\end{equation}
where
\begin{equation} \label{CbarC}
\bar C_\pm^\infty = \mp C_\mp^\infty \, \frac{i (b - a)}{\kappa \mp i m Q} (2B)^{\pm i 2 a}.
\end{equation}
Therefore, the asymptotic behavior of functions ${\cal R}_\pm^\infty = \Sigma^{\mp 1} {\cal \tilde R}_\pm^\infty \simeq (2B)^{\mp i a} z^{\mp i a} {\cal \tilde R}_\pm^\infty$ is
\begin{equation} \label{asymR}
{\cal R}_\pm^\infty \simeq C_\pm^\infty (2B)^{\mp i a} \, z^{\mp i b} + \bar C_\pm^\infty (2B)^{\mp i a} \, z^{\pm i b},
\end{equation}
and the corresponding ingoing and outgoing modes are
\begin{eqnarray} \label{Rz8}
{\cal R}_+^{\infty \mathrm{(in)}} = C_+^\infty (2B)^{- i a} \, z^{- i b}, && \qquad {\cal R}_+^{\infty \mathrm{(out)}} = \bar C_+^\infty (2B)^{- i a} \, z^{i b} = -i \frac{b - a}{\kappa - i m Q} C_-^\infty (2B)^{i a} \, z^{i b},
\nonumber\\
{\cal R}_-^{\infty \mathrm{(out)}} = C_-^\infty (2B)^{i a} \, z^{i b}, && \qquad {\cal R}_-^{\infty \mathrm{(in)}} = \bar C_-^\infty (2B)^{i a} \, z^{- i b} = i \frac{b - a}{\kappa + i m Q} C_+^\infty (2B)^{- i a} z^{- i b}.
\end{eqnarray}
It is obvious that the condition for the existence of propagating modes requires the parameter $b$ must have real value, namely,
\begin{equation} \label{Cond}
(q^2 - m^2) Q^2 - \left( j + \frac12 \right)^2 > 0.
\end{equation}
with half-integer $j$. This condition directly implies $q^2 > m^2$ which is satisfied by known physical spin-1/2 particles, such as electrons and positrons. Moreover, according to the charge and energy conservations, the black holes in the process of pair production will lose its charge more than mass. This consequence ensures the cosmic censorship conjecture.

Similarly, for analyzing the behavior at the horizon, $z = 1$ ($r = B$), the solutions can be transformed, via~(\ref{T1}), to the following expressions
\begin{eqnarray}
{\cal \tilde R}_\pm^H &=& C_\pm^H \, F\left( \pm i (b - a), \mp i (b + a); \frac12 \pm i (\tilde a - a); 1 - z \right)
\nonumber\\
&+& \bar C_\pm^H (z - 1)^{\frac12 \mp i (\tilde a - a)} z^{\frac12 \pm i (\tilde a + a)} \, F\left( 1 \pm i (b + a), 1 \mp i (b - a); \frac32 \mp i (\tilde a - a); 1 - z \right),
\end{eqnarray}
and the coefficients are
\begin{eqnarray} \label{C18}
C_\pm^H &=& \frac{\Gamma(1 \pm i 2b) \Gamma(\frac12 \mp i (\tilde a - a))}{\Gamma(1 \pm i (b + a)) \Gamma(\frac12 \pm i (b - \tilde a))} C_\pm^\infty + \frac{\Gamma(1 \mp i 2 b) \Gamma(\frac12 \mp i (\tilde a - a))}{\Gamma(1 \mp i (b - a)) \Gamma(\frac12 \mp i (b + \tilde a))} \bar C_\pm^\infty,
\nonumber\\
\bar C_\pm^H &=&  \frac{\Gamma(1 \pm i 2b) \Gamma(-\frac12 \pm i (\tilde a - a))}{\Gamma(\pm i (b - a)) \Gamma(\frac12 \pm i (b + \tilde a))} C_\pm^\infty + \frac{\Gamma(1 \mp i 2 b) \Gamma(-\frac12 \pm i (\tilde a - a))}{\Gamma(\mp i (b + a)) \Gamma(\frac12 \mp i (b - \tilde a))} \bar C_\pm^\infty.
\end{eqnarray}
Thus, near horizon the solutions ${\cal R}_\pm^H = \Sigma^{\mp 1} {\cal \tilde R}_\pm^H \simeq (2B)^{\mp i a} (z - 1)^{\pm i \frac{\tilde a - a}2} {\cal \tilde R}_\pm^H$ behave like
\begin{equation}
{\cal R}_\pm^H \simeq C_\pm^H \, (2B)^{\mp i a} (z - 1)^{\pm i \frac{\tilde a - a}2} + \bar C_\pm^H \, (2B)^{\mp i a} (z - 1)^{\frac12 \mp i \frac{\tilde a - a}2}.
\end{equation}
The ingoing and outgoing modes, for case $\tilde a - a > 0$ which covers the extremal limit $B \to 0$ (i.e. $\tilde a \to \infty$), are
\begin{eqnarray} \label{Rz1}
{\cal R}_+^{H \mathrm{(out)}} = C_+^H (2B)^{- i a} \, (z - 1)^{i \frac{\tilde a - a}2}, && \qquad {\cal R}_+^{H \mathrm{(in)}} = \bar C_+^H (2B)^{- i a} \, (z - 1)^{\frac12 - i \frac{\tilde a - a}2} \to 0,
\nonumber\\
{\cal R}_-^{H \mathrm{(in)}} = C_-^H (2B)^{i a} \, (z - 1)^{-i \frac{\tilde a - a}2}, && \qquad {\cal R}_-^{H \mathrm{(out)}} = \bar C_-^H (2B)^{i a} \, (z - 1)^{\frac12 + i \frac{\tilde a - a}2} \to 0.
\end{eqnarray}
Due to the term $(z - 1)^{1/2}$, both modes ${\cal R}_+^{H \mathrm{(in)}}$ and ${\cal R}_-^{H \mathrm{(out)}}$ approach to zero at the horizon.


\section{Spinor Particle Creation}\label{sect:spinorPC}
The pair production process, as discussed in the previous work for scalar field case~\cite{Chen:2012zn}, can be captured by the ratios of the ingoing/outgoing fluxes at the horizon and the asymptotic with appropriate boundary conditions. Actually, there are two equivalent boundary conditions associated to the pair production depending on two complementary particle or antiparticle viewpoints. In this paper, we adopt the particle viewpoint boundary condition which imposes no ingoing flux at the infinity. In such case, the other three non-vanishing fluxes have the following intuitive physical interpretation. The outgoing flux at infinity, $D_\infty^\mathrm{(out)}$, represents the pair produced particles. The outgoing and ingoing flux at horizon, $D_H^\mathrm{(in)}$ and $D_H^\mathrm{(out)}$, correspond to the virtual and re-annihilated particles respectively. The Bogoliubov coefficients, $| {\cal A} |^2$ (vacuum persistence amplitude) and $| {\cal B} |^2$ (mean number of pairs) can be obtained from the following flux ratios~\cite{Kim:2008yt, Kim:2009pg}
\begin{equation}
| {\cal A} |^2 = \frac{| D_H^\mathrm{(out)} |}{| D_H^\mathrm{(in)} |}, \qquad | {\cal B} |^2 = \frac{| D_\infty^\mathrm{(out)} |}{| D_H^\mathrm{(in)} |},
\end{equation}
and the absorption cross section ratio is given by
\begin{equation}
\sigma_\mathrm{abs} = \frac{| D_\infty^\mathrm{(out)} |}{| D_H^\mathrm{(out)} |}.
\end{equation}
However, unlike the scalar field case, the flux conservation for the spinor particle production is indeed $| D_H^\mathrm{(out)} | + | D_\infty^\mathrm{(out)} | = | D_H^\mathrm{(in)} |$, see the discussion for QED in~\cite{Kim:2003qp}, which leads to the Bogoliubov relation $| {\cal A} |^2 + | {\cal B} |^2 = 1$. Accordingly, there is only one independent information in these three physical quantities.

The vector current density of a spinor field is given by
\begin{equation}
J^\mu = \sqrt{-g} \, \bar{\Psi} e_a{}^\mu \gamma^a \Psi,
\end{equation}
in which the Dirac adjoint is defined
\begin{equation}
\bar \Psi = \Psi^\dagger \gamma^0.
\end{equation}
For the general spherically symmetric background it reduces to
\begin{equation}
J^\mu = f^{-\frac12} \Phi^\dagger e_a{}^\mu \gamma^0 \gamma^a \Phi.
\end{equation}
The revelent radial flux, after integrating over whole solid angle $d\Omega = \sin\theta d\theta d\varphi$, is
\begin{eqnarray} \label{Dflux}
D := \int J^r d\Omega &=& - \int (R_+^* \Phi_{\kappa, n}^{+\dagger} + R_-^* \Phi_{\kappa, n}^{-\dagger}) (R_+ \Phi_{\kappa', n'}^- + R_- \Phi_{\kappa', n'}^+) d\Omega
\nonumber\\
&=& - \frac12 ({\cal R}_+^* {\cal R}_+ - {\cal R}_-^* {\cal R}_-).
\end{eqnarray}


By using the flux formula~(\ref{Dflux}), one can straightforwardly compute the ingoing and outgoing fluxes at horizon and asymptotic. The fluxes at the asymptotic region $r \to \infty \; (z \to \infty)$, according to~(\ref{Rz8}), are
\begin{eqnarray}
D_\infty^\mathrm{(in)} &=& - \frac12 \left( | C_+^\infty |^2 - | \bar C_-^\infty |^2 \right) = - \frac{b}{a + b} | C_+^\infty |^2,
\nonumber\\
D_\infty^\mathrm{(out)} &=& - \frac12 \left( | \bar C_+^\infty |^2 - | C_-^\infty |^2 \right) = \frac{b}{a + b} | C_-^\infty |^2.
\end{eqnarray}
Here the relations $| \bar C^\infty_\pm |^2 = \frac{a - b}{a + b} | C^\infty_\mp |^2 $ from equation~(\ref{CbarC}) are used. The particle viewpoint boundary condition $D_\infty^\mathrm{(in)} = 0$ implies $C_+^\infty = 0$, and consequently by~(\ref{CbarC}) implying $\bar C_-^\infty = 0$.

The ingoing and outgoing fluxes at near horizon region, $r \to B \; (z \to 1)$, according to~(\ref{Rz1}) and the boundary condition $C_+^\infty = \bar C_-^\infty = 0$, are
\begin{eqnarray}
D_H^\mathrm{(in)} = \frac12 | C_-^H |^2 &=& \frac12 \left| \frac{\Gamma(1 - i 2b) \Gamma(\frac12 + i (\tilde a - a))}{\Gamma(1 - i (b + a)) \Gamma(\frac12 - i (b - \tilde a))} \right|^2  | C_-^\infty |^2,
\nonumber\\
&=& \frac{\sinh(\pi a + \pi b) \cosh(\pi \tilde a - \pi b)}{\sinh(2 \pi b) \cosh(\pi \tilde a - \pi a)} \, \frac{b}{b + a} | C_-^\infty |^2,
\\
D_H^\mathrm{(out)} = \frac12 | C_+^H |^2 &=& - \frac12 \left| \frac{\Gamma(1 - i 2 b) \Gamma(\frac12 - i (\tilde a - a))}{\Gamma(1 - i (b - a)) \Gamma(\frac12 - i (b + \tilde a))} \right|^2 \frac{b - a}{b + a} \, | C_-^\infty |^2
\nonumber\\
&=& \frac{\sinh(\pi a - \pi b) \cosh(\pi \tilde a + \pi b)}{\sinh(2 \pi b) \cosh(\pi \tilde a - \pi a)} \, \frac{b}{b + a} | C_-^\infty |^2.
\end{eqnarray}
Finally, we obtain the closed form for the Bogoliubov coefficients
\begin{eqnarray} \label{AB_result}
| {\cal A} |^2 &=& \frac{| D_H^\mathrm{(out)} |}{| D_H^\mathrm{(in)} |} = \frac{\sinh(\pi a - \pi b) \cosh(\pi \tilde a + \pi b)}{\sinh(\pi a + \pi b) \cosh(\pi \tilde a - \pi b)},
\nonumber\\
| {\cal B} |^2 &=& \frac{| D_\infty^\mathrm{(out)} |}{| D_H^\mathrm{(in)} |} = \frac{\sinh(2 \pi b) \cosh(\pi \tilde a - \pi a)}{\sinh(\pi a + \pi b) \cosh(\pi \tilde a - \pi b)},
\end{eqnarray}
and the absorption cross section ratio
\begin{equation} \label{sigma_result}
\sigma_\mathrm{abs} = \frac{| D_\infty^\mathrm{(out)} |}{| D_H^\mathrm{(out)} |}= \frac{\sinh(2 \pi b) \cosh(\pi \tilde a - \pi a)}{\sinh(\pi a - \pi b) \cosh(\pi \tilde a + \pi b)}.
\end{equation}
It is worth to note that an interesting relation between the absorption cross section ratio and the mean number of pairs: $\sigma_\mathrm{abs}(b \to -b) = - | {\cal B} |^2$. The vacuum persistence amplitude, mean number of pairs and absorption cross section ratio for spinor and scalar particle productions are greatly analogous. The difference of three parameters $a, \tilde a$ and $b$ is explained after the definition in eq.(\ref{aab}). Moreover, the expressions in eqs.(\ref{AB_result}, \ref{sigma_result}) for spinor particle production can be easily transformed to the results of scalar particle case by $\sinh(\pi a \pm \pi b) \to \cosh(\pi a \pm \pi b)$ and $\cosh(\pi \tilde a - \pi a) \to \sinh(\pi \tilde a - \pi a)$.

\section{Dual CFT Description}\label{sect:holo}
Recall that the spacetime background is the near horizon region of a near extreme RN black hole which is dual to a 2-dimensional CFT with left- and right-hand central charges and temperatures~\cite{Chen:2009ht, Chen:2010bsa, Chen:2010as, Chen:2010yu, Chen:2010ywa, Chen:2011gz}
\begin{equation}
c_L = c_R = \frac{6 Q^3}{\ell}, \qquad T_L = \frac{\ell}{2\pi Q}, \qquad T_R = \frac{\ell B}{\pi Q^2},
\end{equation}
where $\ell$ is a free parameter which can be interpreted as a measure of the U(1) bundle. In addition, from the asymptotic solution~(\ref{Rz8}), one can determine the right-moving conformal dimension of the spinor operator dual to the charged spinor field via $SL(2, R)$ isometry~\cite{Hartman:2009nz}, which is complex
\begin{equation}
h_R = \frac12 \pm i b,
\end{equation}
and the parameters $b$ is defined in~(\ref{aab}). Note that $h_R$ is of the same form as the conformal dimension of the scalar operator dual to the charged scalar field in the same spacetime background although $b$ possesses different value with that in~\cite{Chen:2012zn}. The condition for the existence of propagating mode requires that the value of parameter $b$ should be real, i.e., $(q^2 - m^2) Q^2 - \kappa^2 > 0$. From the field/operator duality in the AdS/CFT correspondence, the conformal dimension of spinor operator in the boundary $d$-dimensional CFT is $\Delta = \frac{d}{2} + |m_{\rm eff}|$, where $m_{\rm eff}$ is the effective mass of the bulk spinor field~\cite{Henningson:1998cd}. The Breitenlohner-Freedman (BF) bound is violated when $m_{\rm eff}$ becomes imaginary. Therefore, the inequality $(q^2 - m^2) Q^2 - \kappa^2 > 0$ just can be rewritten as the violation of BF bound for spinor field in AdS$_2$ (or effectively AdS$_3$) spacetime, i.e. the effective mass square of the spinor field satisfies
\begin{equation}\label{BFspinor}
m^2_{\rm eff} \equiv m^2 - q^2 + \frac{\kappa^2 }{Q^2} \leq 0,
\end{equation}
resulting in instabilities for the bulk spinor fields and dual boundary spinor operators.

The absorption cross section ratio of the spinor field in eq.(\ref{sigma_result}) can be rewritten into a more explicit form as
\begin{equation}\label{abspinor}
\sigma_{\rm abs} = \frac{\sinh(2 \pi b)}{\pi^2 (a - b)} \cosh(\pi a - \pi \tilde{a}) \Bigl| \Gamma\left( 1 + i (b - a) \right) \Bigr|^2 \left| \Gamma\left( \frac12 + i (b + \tilde{a}) \right) \right|^2.
\end{equation}
This version is more convenient to compare with the standard absorption cross section ratio of spinor operators of the dual CFT
\begin{eqnarray} \label{abcft}
\sigma_{\rm abs} &\sim & \frac{(2 \pi T_L)^{2h_L-1}}{\Gamma(2 h_L)} \frac{(2 \pi T_R)^{2 h_R-1}}{\Gamma(2 h_R)} \cosh\left( \frac{\omega_L - q_L \Omega_L}{2 T_L} + \frac{\omega_R - q_R \Omega_R}{2 T_R} \right)
\nonumber\\
&& \times \left| \Gamma\left( h_L + i \frac{\omega_L - q_L \Omega_L}{2 \pi T_L} \right) \right|^2 \left| \Gamma\left( h_R + i \frac{\omega_R - q_R \Omega_R}{2 \pi T_R} \right) \right|^2,
\end{eqnarray}
where ($q_L, q_R$) and ($\Omega_L, \Omega_R$) are the charges and chemical potentials of the left- and right- hands operators, respectively. In addition to the right-moving conformal weight, the left-moving conformal dimension of the spinor operator dual to the spinor field can be identified as
\begin{equation}
h_L = 1 \pm i b,
\end{equation}
which satisfies the natural relation $|h_L - h_R| = \frac 12 = \pm s$, giving the spins of the fermions, e.g., the electron and positron. Similar arguments can be found in the holographic study of (non)-fermion liquids in near extreme RN-AdS black brane~\cite{Liu:2009dm, Iqbal:2009fd, Faulkner:2009wj, Faulkner:2011tm}.

To further compare the results of absorption cross section ratios in eqs.(\ref{abspinor}, \ref{abcft}), recall that there is an identification between the first law of thermodynamics of the black hole and that of the dual CFT, i.e. $\delta S_{\rm BH} = \delta S_{\rm CFT}$, we have
\begin{equation}
\frac{\delta M}{T_H} - \frac{\Omega_H \delta Q}{T_H} = \frac{\tilde{\omega}_L}{T_L} + \frac{\tilde{\omega}_R}{T_R},
\end{equation}
where the black hole Hawking temperature and chemical potential are $T_H = \frac{B}{2 \pi Q^2}, \, \Omega_H = A_\tau(B) = - B/Q$ and the ``total'' energies are $\tilde{\omega}_L = \omega_L - q_L \Omega_L, \, \tilde{\omega}_R = \omega_R - q_R \Omega_R$. Together with the identifications $\delta M = \omega$ and $\delta Q = -q$ (the minus sign corresponds to the convention ``$-q$'' in the operator $D_\mu\equiv \partial_\mu + \Gamma_\mu - i q A_\mu$ for the EoM of the charged spinor field), so we can determine that
\begin{equation}
\tilde{\omega}_L = - q \ell \quad {\rm and} \quad \tilde{\omega}_R = 2 \omega \ell.
\end{equation}
Then one can see that the absorption cross section ratio in eq.(\ref{abspinor}) matches with the CFT's result eq.(\ref{abcft}) only up to some numerical factors. In addition, by the property $\sigma_\mathrm{abs}(b \to -b) = - | {\cal B} |^2$, the mean number of pairs $| {\cal B} |^2$ indeed also matches with the CFT two-point function for fermion~(\ref{abcft}).

In addition, the retarded Green's functions can be obtained from asymptotical behavior of eq.(\ref{asymR}) by adopting the ingoing boundary condition $D_H^\mathrm{(out)} = 0$ at the black hole horizon implying, according to eq.(\ref{Rz1}), $C_+^H = 0$. There are two retarded Green's functions depending on two possible choices of sources, either $G_R^+ \sim {\bar C}^\infty_+ / C^\infty_+, G_R^- \sim C^\infty_- / {\bar C}^\infty_-$ or $\tilde G_R^\pm = 1/G_R^\pm$. According to the relations in eq.(\ref{C18}), the condition $C_+^H = 0$ gives
\begin{equation}
G_R^+ \sim \frac{{\bar C}^\infty_+}{C^\infty_+} = \frac{\Gamma\left(2 i b \right) \Gamma\left( 1- i b + i a \right) \Gamma\left( \frac12 - i b - i \tilde{a} \right)}{\Gamma\left( -2 i b \right) \Gamma\left( 1 + i b + i a \right) \Gamma\left( \frac12 + i b - i \tilde{a} \right)}, \quad G_R^- \sim \frac{C^\infty_-}{{\bar C}^\infty_-} = \frac{a + b}{a - b} G_R^+,
\end{equation}
which indicates that the conformal dimensions of the left- and right-hand spin-$\frac12$ spinor operators are $h_L = 1 - i b$ and $h_R = \frac12 - i b$. While the other type of retarded Green's functions is
\begin{equation}
\tilde G_R^+ = \frac1{G_R^+} \sim \frac{\Gamma\left( -2 i b \right) \Gamma\left( 1 + i b + i a \right) \Gamma\left( \frac12 + i b - i \tilde{a} \right)}{\Gamma\left( 2 i b \right) \Gamma\left( 1 - i b + i a \right) \Gamma\left( \frac12 - i b - i \tilde{a} \right)}, \quad \tilde G_R^- = \frac1{G_R^-} = \frac{a - b}{a + b} \tilde G_R^+,
\end{equation}
which shows that $h_L = 1 + i b$ and $h_R = \frac12 + i b$. These results are also consistent with the result from the dual CFT$_2$ side in which the retarded Green function $G_R(\omega_L, \omega_R)$ is obtained via analytic continuation from the Euclidean correlator (in terms of the Euclidean frequencies $\omega_{EL} = i \omega_L$, and $\omega_{ER} = i \omega_R$)
\begin{eqnarray}\label{CFTGR}
G_E(\omega_{EL}, \omega_{ER}) &\sim& T_L^{2 h_L - 1} T_R^{2 h_R - 1} \mathrm{e}^{i \frac{\tilde\omega_{EL}}{2 T_L}} \mathrm{e}^{i \frac{\tilde\omega_{ER}}{2 T_R}}
\nonumber\\
&& \times \Gamma\left( h_L - \frac{\tilde\omega_{EL}}{2 \pi T_L} \right) \Gamma\left( h_L + \frac{\tilde\omega_{EL}}{2 \pi T_L} \right) \Gamma\left( h_R - \frac{\tilde\omega_{ER}}{2 \pi T_R} \right) \Gamma\left( h_R + \frac{\tilde\omega_{ER}}{2 \pi T_R} \right),
\end{eqnarray}
on the upper half complex $\omega_{L,R}$-plane
\begin{equation}\label{CFTGER}
G_R(i \omega_L, i \omega_R) = G_E(\omega_{EL}, \omega_{ER}), \qquad \omega_{EL}, \omega_{ER} > 0,
\end{equation}
where $\tilde\omega_{EL} = \omega_{EL} - i q_L \mu_L$ and $\tilde\omega_{ER} = \omega_{ER} - i q_R \mu_R$, and the Euclidean frequencies $\omega_{EL}$ and $\omega_{ER}$ take discrete values of the Matsubara frequencies
\begin{equation}\label{Matsubara}
\omega_{EL} = 2 \pi m_L T_L, \qquad \omega_{ER} = 2 \pi m_R T_R,
\end{equation}
and $m_L, m_R$ are half integers here.

Furthermore, the quasinormal modes are determined by the boundary conditions $D_H^\mathrm{(out)} = 0$ (i.e. $C_+^H = 0$ or $\bar{C}_-^H = 0$) and $D_\infty^\mathrm{(in)} = 0$ (i.e. $C^\infty_+ = 0 = \bar{C}^\infty_-$). From eq.(\ref{C18}), the nontrivial solutions satisfying these conditions are
\begin{equation}
\frac1{\Gamma\left( 1 - i b + i a \right) \Gamma\left( \frac12 - i b - i \tilde{a} \right)} = 0, 
\end{equation}
which give
\begin{eqnarray}\label{quasinormal}
\omega_N = - \frac{bB}{Q^2} - i \left( \frac12 + N \right) \frac{B}{Q^2}, \qquad (N = 0, 1, \cdots),
\end{eqnarray}
and they match with the poles of the retarded Green's functions of the dual CFT$_2$, too.

\section{Conclusions and discussions}\label{sect:conclusion}
In this paper, we studied the spinor particle pair production for near extremal RN black holes without backreaction. The Dirac equation was solved in the near horizon region where the spacetime structure is $\mathrm{AdS}_2 \times S^2$ and the background electric field is constant in the radial direction which is effectively a warped AdS$_3$ ($= \mathrm{AdS}_2 \times S^1$) $\times$ $S^2$~\cite{Chen:2010bsa}. The near horizon region contains the causal horizon and dominated electric field which capture both essential contributions: the Hawking radiation and the Schwinger mechanism. Exact spinor solutions were obtained in terms of spherical spinors and hypergeometric functions. Thus one can explicitly compute the ingoing and outgoing fluxes on both inner and outer boundaries. By imposing the particle viewpoint boundary condition, the physical quantities associated to the pair production can be derived by the ratios of boundary fluxes. In particular, the explicit expressions of the vacuum persistence amplitude, the mean number of pairs and the absorption cross section ratio were obtained. Similar to the charged scalar field case, the existence condition of the spinor pair production is actually corresponding to the instability of probe fields in the $\mathrm{AdS}_2$, i.e. violating the BF bound. Moreover, the condition leads to the black holes lose their charge more than mass in the pair production precess. This consequence ensure the cosmic censorship conjecture. The holographic dual CFT description of the spinor particle pair production was also studied in the light of the RN/CFT correspondence. It was showed that both the left- and right-moving conformal dimensions of the spinor operators dual to the bulk spinor field are complex, indicating the spinor operators are unstable. In addition, the CFT fermionic absorption cross section ratio and the retarded Green's function notably agreed with the results computed from the gravity side, and thus the holographic description of the Schwinger pair creation ratio could be understood via its relation with the absorption cross section ratio. Our results revealed further information about the dual CFT$_2$ picture for the near extreme RN black hole. It would be interesting to further study the Schwinger effect in charged black holes in asymptotically AdS$_{d+1}$ spacetime, in which the CFT$_2$ dual to near horizon near extreme geometry is the infrared (IR) one while the dual CFT$_d$ on the asymptotic boundary is the ultraviolet (UV) one. By studying the RG flow between the IR and UV CFTs we then can see how the spontaneous pair creation near the black hole horizon evolve to the boundary and affect the dual CFT$_d$, such as the condensed matter system~\cite{CS2014}.

\section*{Acknowledgement}
The authors thank Rong-Gen Cai, Sang Pyo Kim, Chun-Yen Lin, Jian-Xin Lu for helpful discussions. C.M.C. is grateful to the ITP, CAS for hospitality while the paper was in progress. C.M.C. was supported by the Ministry of Science and Technology of the R.O.C. under the grant MOST 102-2112-M-008-015-MY3 and in part by the National Center of Theoretical Sciences (NCTS). J.R.S. was supported by the NSFC under Grant No. 11205058.

\begin{appendix}

\section{Spherical Spinors} \label{ASS}
The Dirac gamma matrices should satisfy the Clifford algebra, specially in the Minkowski spacetime
\begin{equation}
\left\{ \gamma^a, \gamma^b \right\} = 2 \eta^{ab}.
\end{equation}
There are many representations for gamma matrices and the version used in this paper is
\begin{equation}
\gamma^0 = \left( \begin{array}{cc} -i & 0 \\ 0 & i \end{array} \right), \qquad \gamma^i = \left( \begin{array}{cc} 0 & \sigma^i \\ \sigma^i & 0 \end{array} \right), \quad i = 1, 2, 3,
\end{equation}
in which the Pauli matrices are defined as
\begin{equation}
\sigma^1 = \left( \begin{array}{cc} 0 & 1 \\ 1 & 0 \end{array} \right), \quad \sigma^2 = \left( \begin{array}{cc} 0 & -i \\ i & 0 \end{array} \right), \quad \sigma^3 = \left( \begin{array}{cc} 1 & 0 \\ 0 & -1 \end{array} \right).
\end{equation}
The Pauli matrices have the following relations
\begin{equation}
\sigma^i \sigma^j = \delta^{ij} I + i \epsilon^{ijk} \sigma^k,
\end{equation}
and the operator
\begin{equation}
\pi = \frac12( I + i \sigma^1 + i \sigma^2 + i \sigma^3) = \frac12 \begin{pmatrix} 1 + i & 1 + i \\ - 1 + i & 1 - i \end{pmatrix},
\end{equation}
permutes the Pauli matrices as
\begin{equation}
\pi^{-1} \sigma^1 \pi = \sigma^2, \quad \pi^{-1} \sigma^2 \pi = \sigma^3, \quad \pi^{-1} \sigma^3 \pi = \sigma^1.
\end{equation}

In the curved spacetime, the metric can be encoded in an orthonormal frame via tetrad. However, the frame is not unique. In the following we will firstly introduce the spherical spinors in the Cartesian frame and then transform them into rotation frame which is used in this paper.

\subsection{Cartesian Frame}
In the Cartesian frame, ($\hat{\boldsymbol{e}}_x, \hat{\boldsymbol{e}}_y, \hat{\boldsymbol{e}}_z$), the momentum and angular momentum operators, $\bar{\boldsymbol{P}} = - i \boldsymbol{\nabla}, \bar{\boldsymbol{L}} = \hat{\boldsymbol{r}} \times \bar{\boldsymbol{P}}$, in spherical coordinates have the following components
\begin{eqnarray}
&& \bar P_1 = - i \left( \sin\theta \cos\varphi \partial_r + \frac{\cos\theta \cos\varphi}{r} \partial_\theta - \frac{\sin\varphi}{r \sin\theta} \partial_\varphi \right), \\
&& \bar P_2 = - i \left( \sin\theta \sin\varphi \partial_r + \frac{\cos\theta \sin\varphi}{r} \partial_\theta + \frac{\cos\varphi}{r \sin\theta} \partial_\varphi \right),\nonumber\\
&& \bar P_3 = - i \left( \cos\theta \partial_r - \frac{\sin\theta}{r} \partial_\theta \right),\\
&& \bar L_1 = i \left( \sin\varphi \partial_\theta + \cot\theta \cos\varphi \partial_\varphi \right), \quad \bar L_2 = i \left( -\cos\varphi \partial_\theta + \cot\theta \sin\varphi \partial_\varphi \right), \quad \bar L_3 = - i \partial_\varphi,
\end{eqnarray}

The orthogonal spherical spinors $\bar \Phi^\pm_{\kappa, n}(\theta, \varphi)$ with the value of $\kappa = \mp (j + 1/2)$ are defined as~\cite{Th92}
\begin{equation}
\bar \Phi^+_{\mp (j + 1/2), n} = \begin{pmatrix} i \bar \Psi_{j \mp 1/2}^n \\ 0 \end{pmatrix}, \qquad \bar \Phi^-_{\mp (j + 1/2), n} = \begin{pmatrix}  0 \\ \bar \Psi_{j \pm 1/2}^n \end{pmatrix},
\end{equation}
where $\bar \Psi_{j \pm 1/2}^n(\theta, \varphi)$ are two-component spherical spinors which are defined in terms of the spherical harmonics $Y_l^n(\theta, \varphi)$
\begin{equation}
\bar \Psi_{j - 1/2}^n = \begin{pmatrix} \sqrt{\frac{j + n}{2 j}} Y_{j - 1/2}^{n - 1/2} \\ \sqrt{\frac{j - n}{2 j}} Y_{j - 1/2}^{n + 1/2} \end{pmatrix}, \qquad \bar \Psi_{j + 1/2}^n = \begin{pmatrix} \sqrt{\frac{j - n + 1}{2j + 2}} Y_{j + 1/2}^{n - 1/2} \\ -\sqrt{\frac{j + n + 1}{2j + 2}} Y_{j + 1/2}^{n + 1/2} \end{pmatrix}.
\end{equation}
These spherical spinors are completely specified by the quantum numbers of the angular momentum $j$ ($j = l \pm 1/2$) and its projection $n$ ($-l \le n \le l$) where $l$ is an integer. Moreover, they are eigen-spinors of the operator $I + \boldsymbol{\sigma} \cdot \bar{\boldsymbol{L}}$
\begin{equation}
(I + \boldsymbol{\sigma} \cdot \bar{\boldsymbol{L}}) \, \bar \Psi_{j \mp 1/2}^n = \pm (j + 1/2) \bar \Psi_{j \mp 1/2}^n, 
\end{equation}
where
\begin{equation}
\boldsymbol{\sigma} \cdot \bar{\boldsymbol{L}} = \begin{pmatrix} \bar L_3 & \bar L_1 - i \bar L_2 \\ \bar L_1 + i \bar L_2 & - \bar L_3 \end{pmatrix} = \begin{pmatrix} -i \partial_\varphi & \mathrm{e}^{-i \varphi} (- \partial_\theta + i \cot\theta \partial_\varphi) \\ \mathrm{e}^{i \varphi} (\partial_\theta + i \cot\theta \partial_\varphi) & i \partial_\varphi \end{pmatrix}.
\end{equation}
Moreover, the spherical spinors also satisfy the relation
\begin{equation}
\boldsymbol{\sigma} \cdot \hat{\boldsymbol{r}} \, \bar \Psi_{j \mp 1/2}^n = \bar \Psi_{j \pm 1/2}^n,
\end{equation}
where
\begin{equation}
\boldsymbol{\sigma} \cdot \hat{\boldsymbol{r}} = \sin\theta \cos\varphi \sigma^1 + \sin\theta\sin\varphi \sigma^2 + \cos\theta \sigma^3 = \begin{pmatrix} \cos\theta & \sin\theta \mathrm{e}^{- i \varphi} \\ \sin\theta \mathrm{e}^{i \varphi} & -\cos\theta \end{pmatrix}.
\end{equation}
In the Dirac equation in flat spacetime, the spatial derivative is basically $\boldsymbol{\sigma} \cdot \bar{\boldsymbol{P}}$ which can be expressed in terms of $\boldsymbol{\sigma} \cdot \hat{\boldsymbol{r}}$ and $\boldsymbol{\sigma} \cdot \bar{\boldsymbol{L}}$ via the identity
\begin{equation}
\boldsymbol{\sigma} \cdot \bar{\boldsymbol{P}} = -i (\boldsymbol{\sigma} \cdot \hat{\boldsymbol{r}}) \partial_r + \frac{i}{r} (\boldsymbol{\sigma} \cdot \hat{\boldsymbol{r}}) (\boldsymbol{\sigma} \cdot \bar{\boldsymbol{L}}).
\end{equation}
Therefore, the spherical spinors are suitable basis to expand the general solution.

%

\subsection{Rotation Frame}
The other convenient frame for solving the Dirac equation is the rotation frame ($\hat{\boldsymbol{e}}_r$, $\hat{\boldsymbol{e}}_\theta$, $\hat{\boldsymbol{e}}_\varphi$).
For separating the Dirac equation, the analog two-component spherical spinors in rotation frame, $\Psi_{j \pm 1/2}^n$, must have a nice property with respect to the associated angular operator
\begin{equation}
\boldsymbol{K} = \sigma^2 \partial_\theta + \sigma^3 \frac{\partial_\varphi}{\sin\theta} = \left( \begin{array}{cc} \frac{\partial_\varphi}{\sin\theta} & - i \partial_\theta \\ i \partial_\theta & - \frac{\partial_\varphi}{\sin\theta} \end{array} \right).
\end{equation}
Basically the two-component spherical spinors in the rotation and Cartesian frames should be related a similarity transformation $\Psi_{j \pm 1/2}^n = \boldsymbol{S}^{-1} \bar \Psi_{j \pm 1/2}^n$. The similarity transformation~\cite{Villalba:1994mv} turns out to be
\begin{equation}
\boldsymbol{S} = \frac1{\sqrt{\sin\theta}} \, \mathrm{e}^{-i \frac{\varphi}2 \sigma^3} \, \mathrm{e}^{-i \frac{\theta}2 \sigma^2} \, \pi,
\end{equation}
which transforms the operator $\boldsymbol{K}$ as
\begin{equation}
\boldsymbol{S} \boldsymbol{K} \boldsymbol{S}^{-1} = - (\boldsymbol{\sigma} \cdot \hat{\boldsymbol{r}}) (I + \boldsymbol{\sigma} \cdot \boldsymbol{L}).
\end{equation}
One can straightforwardly verify that the spherical spinors satisfy
\begin{eqnarray}
\boldsymbol{K} \Psi_{j \mp 1/2}^n = \mp (j + 1/2) \Psi_{j \pm 1/2}^n,
\end{eqnarray}
and
\begin{equation}
\sigma^1 \Psi_{j \mp 1/2}^n
= \Psi_{j \pm 1/2}^n.
\end{equation}
Therefore, the spherical Dirac spinor in the rotation frame is
\begin{equation}
\Phi^\pm_{\kappa, n} = \boldsymbol{\cal S}^{-1} \bar \Phi^\pm_{\kappa, n}, \qquad \boldsymbol{\cal S} = \begin{pmatrix} \boldsymbol{S} & 0 \\ 0 & \boldsymbol{S} \end{pmatrix}.
\end{equation}
which satisfy the following relations
\begin{equation} \label{g01K}
\gamma^0 \Phi^\pm_{\kappa, n} = \mp i \Phi^\pm_{\kappa, n}, \qquad \gamma^1 \Phi^\pm_{\kappa, n} = \pm i \Phi^\mp_{\kappa, n}, \qquad \boldsymbol{\cal K} \Phi^\pm_{\kappa, n} = i \kappa \Phi^\mp_{\kappa, n},
\end{equation}
where
\begin{equation}
\boldsymbol{\cal K} = \gamma^2 \partial_\theta + \gamma^3 \frac{\partial_\varphi}{\sin\theta} = \begin{pmatrix} 0 & \boldsymbol{K} \\ \boldsymbol{K} & 0 \end{pmatrix}.
\end{equation}


\section{Hypergeometric Functions} \label{AHF}
The hypergeometric equation
\begin{equation} \label{hgeq}
z (1 - z) \partial_z^2 w + [\gamma - (\alpha + \beta + 1) z] \partial_z w - \alpha \beta w = 0,
\end{equation}
has two independent solutions which can be expressed around three regular singular points, i.e. $z = 0, z = 1$ and $z = \infty$, as
\begin{eqnarray}
w &=& a_1 \, F(\alpha, \beta; \gamma; z) + a_2 \, z^{1 - \gamma} F(\alpha - \gamma + 1, \beta - \gamma + 1; 2 - \gamma; z),
\label{Fz0} \\
&=& b_1 F(\alpha, \beta; \alpha + \beta - \gamma + 1; 1 \!-\! z) + b_2 (1 \!-\! z)^{\gamma - \alpha - \beta} F(\gamma - \alpha, \gamma - \beta; \gamma - \alpha - \beta + 1; 1 \!-\! z),
\label{Fz1} \\
&=& c_1 \, z^{-\alpha} F\left( \alpha, \alpha - \gamma + 1; \alpha - \beta + 1; \frac1{z} \right) + c_2 \, z^{-\beta} F\left( \beta, \beta - \gamma + 1; \beta - \alpha + 1; \frac1{z} \right).
\label{Fz8}
\end{eqnarray}

There are a number of mathematical properties for $F(\alpha, \beta; \gamma; z)$, in particular the following ones are useful for our analysis: (i) transformation formula
\begin{eqnarray}
F(\alpha, \beta; \gamma; z)
&=& \frac{\Gamma(\gamma) \Gamma(\gamma - \alpha - \beta)}{\Gamma(\gamma - \alpha) \Gamma(\gamma - \beta)} z^{-\alpha} F\left( \alpha, \alpha - \gamma + 1; \alpha + \beta - \gamma + 1; 1 - \frac1{z} \right)
\\
&& + \frac{\Gamma(\gamma) \Gamma(\alpha + \beta - \gamma)}{\Gamma(\alpha) \Gamma(\beta)} (1 - z)^{\gamma - \alpha - \beta} z^{\alpha - \gamma} F\left( \gamma - \alpha, 1 - \alpha; \gamma - \alpha - \beta + 1; 1 - \frac1{z} \right), \label{T1} \nonumber
\end{eqnarray}
(ii) special values
\begin{equation}
F(\alpha, \beta; \gamma; 0) = 1, \qquad F(\alpha, \beta; \gamma; 1) = \frac{\Gamma(\gamma) \Gamma(\gamma - \alpha - \beta)}{\Gamma(\gamma - \alpha) \Gamma(\gamma - \beta)},
\end{equation}
and (iii) differential formula
\begin{eqnarray}
&& \partial_z F(\alpha, \beta; \gamma; z) = \frac{\alpha \beta}{\gamma} F(\alpha + 1, \beta + 1; \gamma + 1; z),
\\
&& \partial_z \left[ z^\alpha F(\alpha, \beta; \gamma; z) \right] = \alpha z^{\alpha - 1} F(\alpha + 1, \beta; \gamma; z).
\\
&& \partial_z \left[ z^{\gamma -1} F(\alpha, \beta; \gamma; z) \right] = (\gamma - 1) z^{\gamma - 2} F(\alpha, \beta; \gamma - 1; z).
\end{eqnarray}
In addition, the following properties of Gamma function are also needed in our computation
\begin{eqnarray}
&& \Gamma(\alpha + 1) = \alpha \Gamma(\alpha), \qquad \Gamma(\alpha) \Gamma(1 - \alpha) = \frac{\pi}{\sin(\alpha \pi)},
\\
&& \left| \Gamma\left( \frac12 + i y \right) \right|^2 = \frac{\pi}{\cosh(\pi y)}, \quad \left| \Gamma\left( 1 + i y \right) \right|^2 = \frac{\pi y}{\sinh(\pi y)}, \quad \left| \Gamma\left( i y \right) \right|^2 = \frac{\pi}{y \sinh(\pi y)}.
\end{eqnarray}

\end{appendix}


\end{document}